\documentclass{aastex631}

\usepackage{indentfirst}
\usepackage{gensymb}
\usepackage{url}

\graphicspath{{./}{images/}}

\shorttitle{Dracula's Chivito}
\shortauthors{Berghea et al.}

\begin{document}

\title{Dracula's Chivito: discovery of a large edge-on protoplanetary disk with Pan-STARRS}

\author{Ciprian T. Berghea}
\affiliation{U.S. Naval Observatory (USNO), 3450 Massachusetts Avenue NW, Washington, DC 20392, USA}

\author{Ammar Bayyari}
\affiliation{Department of Physics and Astronomy, University of Hawaii, Honolulu, HI 96822, USA}

\author{Michael L. Sitko}
\affiliation{Department of Physics, University of Cincinnati, Cincinnati, OH 45221, USA}
\affiliation{Center for Extrasolar Planetary Systems, Space Science Institute, 4750 Walnut Street, Suite 205, Boulder, CO 80301, USA}

\author{Jeremy J. Drake}
\affiliation{Lockheed Martin, 3251 Hanover Street, Palo Alto, CA94304, USA}

\author{Ana Mosquera}
\affiliation{U.S. Naval Observatory (USNO), 3450 Massachusetts Avenue NW, Washington, DC 20392, USA}
\affiliation{Federated-IT, 1200 G Street NW, Suite 800, Washington DC 20005, USA}

\author{Cecilia Garraffo}
\affiliation{Center for Astrophysics, Harvard \& Smithsonian, 60 Garden Street, Cambridge, MA 02138, USA}

\author{Thomas Petit}
\affiliation{Amateur Astronomer, Kischova 1733, 14000, Praha, Czech Republic}
\affiliation{2SPOT, Chemin du Lac, 38690, Chabons, France}
\affiliation{APO Team, 67150, Erstein-Kratf, France}

\author{Ray W. Russell}
\affiliation{The Aerospace Corporation, Los Angeles, CA, USA}

\author{Korash D. Assani}
\affiliation{Department of Astronomy, University of Virginia, Charlottesville, VA 22904, USA}

\email{tberghea@yahoo.com}


\begin{abstract}

We report the serendipitous discovery of a large edge-on protoplanetary disk in Pan-STARRS (PS1) images. PS1 has five broadband filters designated as g$_{P1}$, r$_{P1}$, i$_{P1}$, z$_{P1}$, y$_{P1}$ (hereafter $grizy$) with mean wavelengths 4866, 6215, 7545, 8679 and 9633 \AA,  respectively. The disk's apparent size in the PS1 images is $\approx11\arcsec$, making this one of the largest known disks on the sky. It is likely a young system, still surrounded by the envelope which is very faint but still visible in the PS1 images in the northern part (alternatively this structure could be filaments from the disk itself). We use the PS1 magnitudes and other available photometric data to construct the spectral energy distribution (SED) of the disk. An  optical spectrum indicates that the obscured star is hot, most likely of type late A. We adopt a distance of 300 pc for this object based on {\it Gaia} DR3 extinctions. We model the system using the HOCHUNK3D radiative transfer software and find that the system is consistent with a hot star of effective temperature 8000~K surrounded by a disk of size 1650 AU and mass 0.2 M$_\odot$ at inclination 82$\degree$.

\end{abstract}


\section{Introduction} \label{sec:intro}

Planet formation within protoplanetary disks around ostensibly similar stars results in a surprisingly large diversity of exoplanetary systems \citep[e.g.,][and references therein]{Winn2015}. The processes responsible involve the hierarchical assemblage of successively larger bodies, beginning with sub-micron-sized dust particles, but the details of these processes and how they lead to the diverse exoplanet demographics revealed by the intensive planet hunting endeavours over the last two decades remain poorly understood. Key to this is dust growth and transport, and the extent to which the dust couples to the gas as a function of grain size, gas temperature and turbulence \citep[][]{ Goldreich1973,Weidenschilling1977,Voelk1980,Dullemond2005,Binkert2023}.

Extensive studies of disks aimed at making progress in this direction have tended to focus on low-inclination objects in which remarkable structures such as spirals, rings and gaps have been resolved \citep[e.g.,][]{ALMAPartnership2015,Ruge2016,Andrews2020}. However, it is difficult to determine the vertical structure of these disks \citep[e.g.,][]{Pinte2016}, the relative distributions of gas and dust, and the radial dependence and extent of dust settling and disk flaring. 
It has been increasingly recognised in recent years that edge-on protoplanetary disks instead offer the opportunity to probe the three-dimensional morphology of disks, sacrificing disk plane details but providing a direct line of sight of the vertical axis. 
The shadows cast by the midplane concentration of dust and gas in highly-inclined disks allow direct observations of scattered light from the central star, unveiling the vertical distribution of dust particles, providing information about their sizes, compositions, and settling mechanisms \citep[][]{Dutrey2017,Villenave2020,wolff2021,Angelo2023}. This different perspective into disk structure and evolution renders nearby edge-on disks in which spatial resolution is maximized of particularly great value.
In this {\em Letter}, we report the serendipitous discovery of a remarkable new edge-on protoplanetary disk candidate using images taken by the Panoramic Survey Telescope and Rapid Response System (Pan-STARRS, hereafter PS1). We shall argue that the central star is an obscured object of intermediate mass and likely a Herbig Ae (HAeBes) star, and that distance constraints imply that the disk is possibly the largest protoplanetary disk found to date.

The structure of the paper is as follows. Section~\ref{sec:discovery} describes the discovery and the PS1 image of the disk. 
In Section \ref{sec:data} we present the photometric data used to construct the SED and the spectroscopic data  to constrain the host star properties. In Section \ref{sec:rt} we model the disk using the HOCHUNK3D radiative transfer code. In Section \ref{sec:dist} we estimate the distance to this object. In Section \ref{sec:discussion} we discuss our results and in Section \ref{sec:conclusion} we summarize our conclusions. 

\section{Discovery of DraChi}
\label{sec:discovery}

\begin{figure*}
\includegraphics[width=0.85\textwidth]{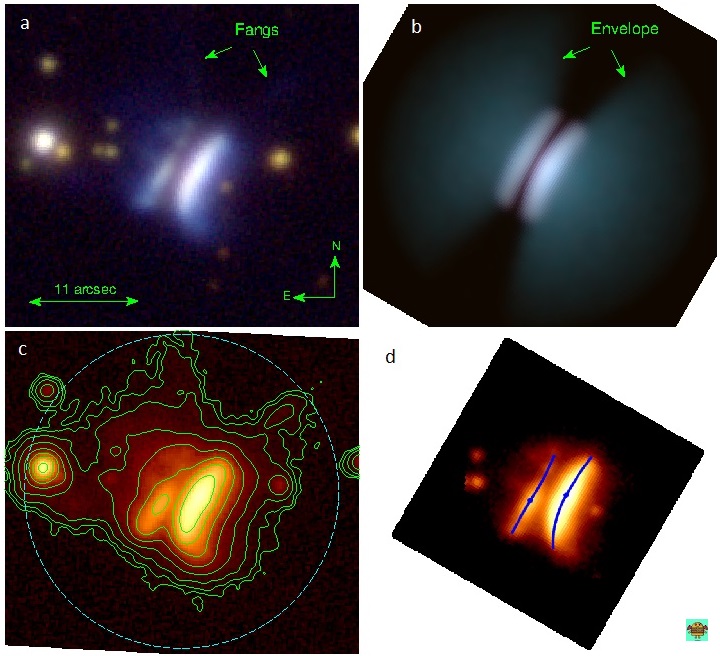}
\caption{a) Color (giy bands) PS1 image of DraChi. The coordinates of the center of the disk are $RA=347.432214^{\circ}$, $DEC=67.394451^{\circ}$. In the northern part of the disk, very faint filaments are seen extending $17\arcsec$ out from both edges of the disk. In the south these features which we call ``fangs'' in the text are likely not visible because of increased extinction. b) The best radiative transfer model image in the same bands. The model image was convolved with the PS1 PSF. The ``edges'' of the envelope are very similar to the faint fangs in panel a. c) PS1 r-band image with contours to show the faint fangs. The blue circle has a radius of $16\arcsec$. d) PS1 r-band image showing the analysis of the scattered light. The blue lines are spines fitted to the two lobes as described in the text and the peaks of the emission are marked with a dot. A significant lateral asymmetry can be clearly seen, suggesting a tilted inner region.}
\label{fig:image}
\end{figure*}

This remarkable disk was discovered in PS1 images while working on a variability study of active galactic nuclei candidates from \citet{secrest2015} in conjunction with data from the USNO Robotic Astrometric Telescope \citep{zacharias2015}. PS1 observed the entire sky north of $-30^{\circ}$ declination between 2010 and 2014 with exposure times between 30 and 60 seconds. The PS1 telescope has a diameter of 1.8 m, a 7.7 deg$^2$ field of view and a pixel size of 0.258$\arcsec$ \citep{cha2016}. Individual observations were combined in so-called stacks to obtain deeper, higher signal-to-noise images \citep{wat2020}. A composite color PS1 image of the disk is illustrated in panel a of Figure~\ref{fig:image}. The disk has the characteristic bipolar appearance of edge-on disks with the central star being completely obscured  DraChi appears blue in comparison with the stars in the vicinity suggesting a hot star behind the disk. 

The scattered light covers about $11\arcsec$ on the sky, with very faint filaments in the disk's northern part as extensions of both lobes out to $\sim$ 17$\arcsec$, better seen in panel c of Figure \ref{fig:image}, which shows the fainter emission as contours. While these could represent the fainter outer parts of the disk, we think they actually trace a dissipating envelope seen in panel b of Figure \ref{fig:image}, which shows an image obtained from the modeling presented later in Section \ref{sec:rt}. As explained below (Section \ref{sec:discussion}) this suggests the disk is in transition between Class I and II. The left lobe is approximately 5 times fainter than the right lobe (peak values) , suggesting that the disk is not seen perfectly edge-on. The left lobe is darker in the northern part and filament-like features can be seen extending perpendicularly from this part of the disk and also seen as an extended emission in the contours of panel c, suggesting winds and higher extinction. These features can also possibly be part of a filamentary envelope such as seen in IRAS 04302+2247 \citep{Villenave2024}. Finally, the lower edge of the bright lobe shows pronounced flaring, similar to the ``wings'' seen in PDS 144N \citep{perr2006}.

We investigate a possible lateral asymmetry following the method described by \citet{Villenave2020} and \citet{duch2024}, to define a spine for each lobe of the disk, setting the outer edge of each lobe as the limit where the flux drops to 5\% of the peak. \citet{Villenave2024} found such asymmetries in the majority of the edge-on disks investigated and interpret this effect by a tilted inner region of the disk. The result of our analysis is presented in panel d of Figure \ref{fig:image}, which shows a clear lateral offset between the lobes. Some of the offset can be due to the left lobe being fainter in the northern part. The peak emission is marked with a dot in this image. The offsets are 0.8$\arcsec$, 2.3$\arcsec$ and 1.7$\arcsec$ for the lower edge, peak and upper edge, respectively. We note that because the left lobe is fainter in the northern part (extincted?), the peak of this lobe is clearly offset from the center, thus increasing the offset artificially (this is probably also true for the upper edge).

The object was not unknown: it is a bright infrared source (IRAS 23077+6707) and  it was observed at radio wavelengths as a possible pre-main sequence star (PMS) based on the IRAS colors by \cite{wouterloot1993}. Their sample of candidates was selected from IRAS sources based on their colors. It was also identified as a young stellar object candidate in the AKARI Far-Infrared Surveyor catalog of \citet{toth2014} based on the AKARI and WISE colors and as an AGN in the catalog of \citet{secrest2015} based on WISE colors as we already mentioned above. More recently, an optical spectrum was obtained by a French amateur astronomy group looking for planetary nebula (PNe) candidates\footnote{https://planetarynebulae.net.} \citep{ledu2022}. Our object was identified in the PS1 images as an extended blue source, was named Kn~32 and is included in Table~4 of this paper. However the object was never investigated in particular. In addition, photometric data is available from several all-sky surveys as shown in the following section and Table \ref{tbl:photometry}.

There are very few edge-on disks known so far, making this one particularly valuable for study and future observations since it is also very large. \citet{Angelo2023} list a sample of 22 T Tauri edge-on disks. The only previously known HAeBe edge-on systems we are aware of are the well-known Gomez's Hamburger \citep[``GoHam'',][] {ruiz1987} and PDS 144N \citep{perr2006}. DraChi is by far the largest of all these, and at $\sim11\arcsec$ in angular extent is the largest edge-on protoplanetary disk discovered to date. The well known GoHam is $\sim$ $8\arcsec$ in size and just as DraChi is not associated with any known star forming region. Submillimeter Array observations revealed Keplerian rotation about the central star and established GoHam as a massive circumstellar disk \citep{buj2008,wood2008}. 
Since our disk is similar to GoHam, and in addition has faint filaments (``fangs")  extending far out from the northern edges (see Figure \ref{fig:image}), we name it ``Dracula's Chivito"\footnote{One of the authors grew up in Transylvania, not far from the place where Vlad (Draculea) Tepes lived, and another author is from Uruguay, where ``chivito" is a traditional sandwich and it is similar to a hamburger.}, DraChi hereafter.

\section{Photometry and Spectroscopy}\label{sec:data}

Stacked PS1 images were retrieved from the STScI archive \citep{ps1dr2} for the $grizy$ filters centered on the sky location $RA=347.4322139^{\circ}$, $DEC=67.39445091^{\circ}$. 
To perform the photometric analysis, we followed the procedure described in the PS1 archive\footnote{https://outerspace.stsci.edu/display/PANSTARRS/PS1+Stack+images} and by \citet{wat2020}, using an aperture of 14$\arcsec$ to cover the whole disk. There is a small amount of contamination from at least one faint star on the right side of the disk, but at the level of 1.1\% it is insignificant for the purpose of our analysis. 

\begin{figure}[ht!]
\begin{center}
\includegraphics[width=0.7\textwidth]
{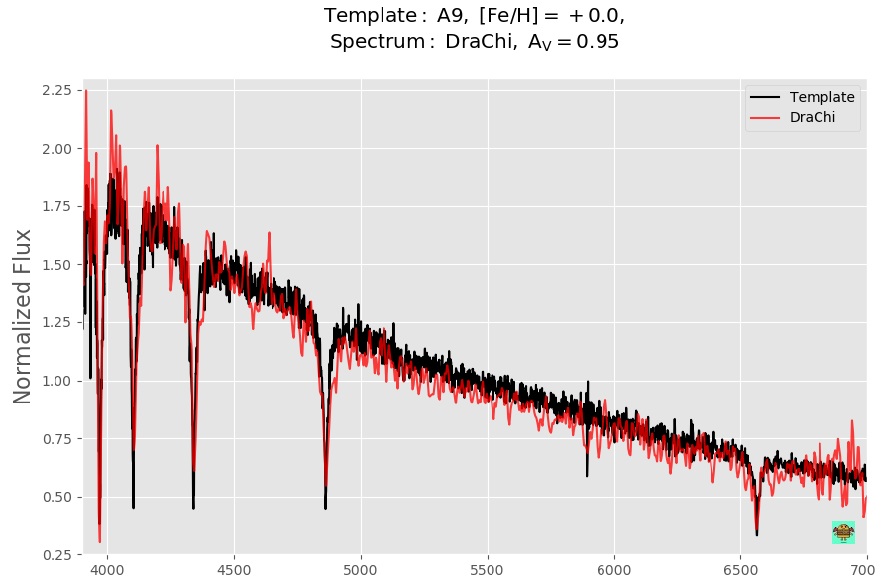}
\end{center}
\caption{Template best fit to the optical spectrum of DraChi. The spectrum was de-reddened by $A_V = 0.95$.}
\label{fig:pyhammer}
\end{figure}

We also searched for all the available archival photometric data on DraChi. These and our own photometry are shown in Table~\ref{tbl:photometry}. As noted for PS1, some contamination from stars is expected in all these measurements, but it is likely very small, especially at longer wavelengths, where the disk dominates the nearby sources. 2MASS and ALLWISE both classify DraChi as an extended source and provide photometry for several apertures with different sizes. The ALLWISE  apertures between 8.25$\arcsec$ and 13.75$\arcsec$ radius vary by only 0.2 mag and the profile, corrected and elliptical magnitudes \citep{cutri2013} of this source also vary by only 0.2 mag. Similar variations are shown by different apertures in the 2MASS extended source catalog - Kron, isophotal, fit extrapolation \citep{jarr2000} are within 0.1 mag, while the fixed radii apertures between 7$\arcsec$ and 15$\arcsec$ also vary by less than 0.1 mag. Finally, the values measured by different instruments in Table~\ref{tbl:photometry} match very well with each other, for example the 12 $\mu$m value is almost the same for ALLWISE and IRAS. We estimate therefore that a conservative contamination level is always less than 25\%. We also note that some variability at this level is also expected for PMS stars \citep[e.g.][]{esp2011}. We use these data in Section \ref{sec:rt} to construct the SED and constrain the simulated models.

The optical spectrum obtained by the French amateur astronomers was taken on Aug 3, 2019. More information about the exemplary work of this group can be found in \citet{ledu2022}. The spectrum was taken with a Celestron C8 Edge H with a 23~ $\mu$m slit with a low resolution of R = 600. The spectrum was dark, bias, and flat-field corrected. The instrumental response was derived from the reference spectrum of HD 219485 acquired under the same observing conditions as the target. Wavelength calibration was done via an argon-neon calibration lamp, and was checked against absorption lines of the reference star in the blue part of the spectrum. The final spectrum was produced after removal of atmospheric lines and cosmic rays. More details of the observation can be found on their website\footnote{\url{ https://planetarynebulae.net/EN/tableau\_spectres.php}}. 

The optical emission from edge-on disks is produced by scattering in the disk since the star is completely obscured. Optical spectra sometimes show different features compared to the star emission such as veiling related to accretion and emission lines from winds and jets. However, the photospheric absorption spectrum agrees well with that of the hidden star and therefore can be used to estimate its spectral type \citep{app2005, flores2021}.

We fitted the spectrum to star templates based on spectra taken by the Sloan Digital Sky Survey using the software PyHammer v2.0 \citep[\url{https://github.com/BU-hammerTeam/PyHammer}, ][]{kess2017, roul2020}.  PyHammher finds the best fit automatically by comparing important spectral lines to the template spectra. Extinction is not taken into account by PyHammer so we corrected our spectrum before the fit to match the slope of the continuum. find a best fit for a type A9 star and absorption A$_V$ = 0.95 (see Figure~\ref{fig:pyhammer}). An A9 star on the main sequence has a mass of approximately 1.7 M$_{\odot}$ \citep{cox2000}. Given that many parameters of the hidden star are unknown such as the age, a conservative mass range for the star in DraChi is 1.5 - 2.0 M$_{\odot}$. For masses in this range, models for PMS stars \citep{palla1999} imply effective temperatures of 6500 - 8500 K.

\section{Radiative Transfer Modeling}\label{sec:rt}

To investigate very approximate disk parameters based on the various data at hand we use the three-dimensional dust continuum Monte-Carlo radiative transfer code HOCHUNK3D\footnote{https://gemelli.colorado.edu/~bwhitney/codes/codes.html} \citep{whitney2013}. The basic structure of a typical flared disk can be described by the height of the disk, the surface density, and the density structure, respectively:
\begin{equation}
H(r) = H_0 \left( \frac{r}{r_\mathrm{pivot}} \right)^{\beta}, 
\end{equation}
\begin{equation}
\Sigma(r) = \Sigma_0 \times ({r \, / \, r_\mathrm{pivot}})^{-(\alpha - \beta)}, 
\end{equation}
\begin{equation}
\rho(r,z) = {\rho_0 \over { ({r\, / \, r_\mathrm{pivot}})^{\alpha} } } \exp[{-{1 \over 2} ({z\, /\, H(r)})^{2}}]
\end{equation}
where $\beta$ is the disk flaring exponent, $\alpha$ is the disk radial density exponent, and $H_0$ and $\Sigma_0$ are the scale height and surface density of the disk at the pivot radius, $r_\mathrm{pivot}$.

HOCHUNK3D creates both images and SEDs. We do not use the images for model fitting, the models are only fit to the photometric data presented in Sect. \ref{sec:data}. We do use the PS1 images however and also the optical spectrum to impose constraints on the model which helps to reduce the parameter space of the model which is quite large and to eliminate some of the degeneracy between parameters.  Assuming the star type A9 obtained from the optical spectrum we limit the star temperature to the range 6500 - 8500~K. The PS1 images show a peak flux factor difference between the lobes of approximately 5. Assuming this is caused by the inclination we constrain the inclination angle to 79 - 84$\degree$. Using a toy HOCHUNK3D model we find that these angles correspond to flux factors of approximately 10 - 3 between the lobes. Similarly, the distance between the lobes can be used to constrain the scale height for a given distance to DraChi. We adopt a distance of 300 pc based on the arguments presented in Sect. \ref{sec:dist}. The PS1 images show that the distance between lobes (see also Figure \ref{fig:image}) is 3.2$\arcsec$.  We constrain the scale height to 25 - 50 at a pivot radius of 500~AU, which corresponds to distances between the lobes of 2.5$\arcsec$ - 3.8$\arcsec$. 

\citet{miotello2014} describe Class I objects as disks with partially dissipating envelopes that contain large grains embedded deep within the inner regions of the envelopes. Our HOCHUNK3D consists of the A9 star, two coplanar disks, one with small dust grains and one with larger grains to simulate grain settling and an Ulrich envelope \citep{ulr1976}. The star mass and radius were set at 1.7 $M_\odot$ and 2.0 $R_\odot$. Both disks begin at the sublimation radius of 0.25 AU which is the dust destruction radius as defined in \citet{rob2006}. This is calculated for  approximate values of $R_{star} = 2.0~R_{\odot}$ and $T_{star} = 8000~K$ and dust sublimation temperature $T_{sub} = 1600~K$ as $R_{sub} = R_{star}(T_{sub}/T_{star})^{-2.1}  = 51~R_{\odot}$. The small grain disk ends at 1650 AU which correspond to 11$\arcsec$ at 300 pc, whereas the large grain disk ends at 500 AU which best fits the near-IR photometry. We note that the geometric constrains we imposed above only apply to the small grains disk because the large grains disk does not affect the model image significantly. The Ulrich envelope has an ambient density at the inner sublimation radius of $1.5 \times 10^{-21}$~g/s and extends to 3000~AU. These were chosen so that the envelope in the model image is faint and matches the extent of the fangs, the envelope component did not significantly change the model SED.

\begin{figure}
\begin{center}
\includegraphics[width=0.6\textwidth]
{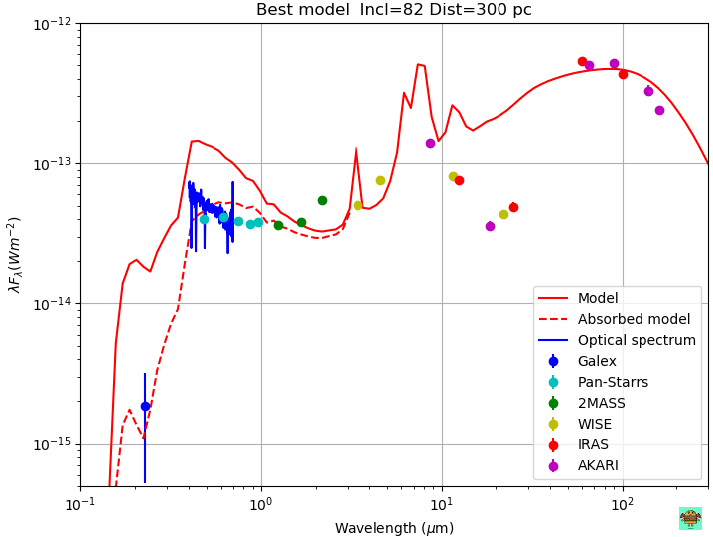}
\end{center}
\caption{SED constructed with all the data presented in this paper and our best HOCHUNK3D model. We also show the extincted model with A$_V =  0.95$ (dashed line). The optical spectrum was scaled to match the PS1 r band. Most error bars (formal errors) are too small to be visible.}
\label{fig:sed}
\end{figure}

For the dust composition we follow \citet[][for GoHam]{wood2008} and \citet[][for HAeBe star HD 163296]{Pik2021} who used Model 1 from \citet{wood2002}. As \citet{wood2008} pointed out, millimeter observations can provide the slope of the dust opacity and thus constrain the dust size distribution, but given our limited data we adopt this model which has been used successfully for other edge-on disks as in the works quoted. This model extends the standard ISM distribution from \citet[][KMH model]{kim1994} to include larger grains by combining a power-law with exponential cutoff for components of amorphous carbon and astronomical silicates with solar abundance constraints for both, and a max radius of a$_{max}$ = 1000 $\mu$m. We therefore used this model for the large grain disk while for the small grain disk the standard KHM model was used, where the size distribution is determined by maximum-entropy fitting of the interstellar extinction curve. For this model the max grain sizes are a$_{max}$ = 0.16 $\mu$m for silicates and a$_{max}$ = 0.28 $\mu$m  for graphite. Both disks also include polycyclic aromatic hydrocarbons and the very small grains (PAHs/VSGs) dust model from \cite{draine2007}. Hot stars can produce significant PAH emission \citep{perr2006} and we found that including this component improved the fit in the near-IR. The Ulrich envelope has the same composition as the small grains disk.

Exhaustive radiative transfer modelling is beyond the scope of this letter; the intention here is instead to estimate a plausible set of coarse disk parameters that explain the data reasonably well. Given the constraints imposed by  the PS1 image and the A9 spectral type presented above, we explored a range of parameters to find a model that best fits the SED and especially the far-IR as it is not affected by extinction. The parameters ranges and best fits are: inclination (79 - 84, 82$\degree$), star temperature (6500 - 8500, 8000 K), disk mass (0.01 - 0.5, 0.2 $M_\odot$), fraction of disk mass in large grains (0.1 - 0.5, 0.2), small grains disk scale height at 500 AU (25 - 50, 30 AU), large grains disk scale height (1 - 20, 5 AU), flaring exponent $\beta$ for the small grains disk (1.0 - 1.5, 1.15) for the large grains disk (1.0 - 1.5, 1.4), radial density exponent $\alpha$ for both the small and large grains disk (1 - 2.5, 2.2). The dust-to-gas ratio was taken at the standard fraction of 0.01. The star luminosity was adjusted to fit the far-IR and then extinction was added to fit the optical and near-IR.

The best model is presented in panel b of Figure~\ref{fig:image} and Figure \ref{fig:sed} which show the model image convolved with the PS1 PSF, and the model SED, respectively. The star luminosity 11.5 L$_{\odot}$. A line-of-sight absorption A$_V = 0.95$ was added to the model to match the optical and near-IR. We emphasise that this model is not unique, but it provides a reasonable match to the available data. However, there are significant discrepancies in the near and mid-IR range. It is also remarkable how well the envelope included in the model resembles the faint ``fangs'' seen in the PS1 image in the northern part of the disk (Figure \ref{fig:image}). 

\section{Distance}
\label{sec:dist}

Since DraChi is not associated with any known star-forming region the distance cannot be precisely determined, similar to GoHam which has estimates ranging from 250 to 500~pc or PDS 144N with an even wider range of 140 to 2000 pc \citep{perr2006}. 
However, the galactic coordinates of DraChi (l = 113.4$\degree$, b = 6.4 $\degree$) places it close to the Cepheus Flare Shell. This is thought to be an expanding supernova bubble and is home to a complex of molecular clouds and star-forming regions. It is located at l = 120$\degree$ and b = 17$\degree$, a distance of 300 pc and is 9.5$\degree$ in radius. Recently, {\it Gaia} data has been used to measure the distances to some of the dark clouds and young stars in this region  \citep{sharma2022, szi2021} with values ranging from 150 to 900 pc, with the majority of stars between 330 and 370 pc. These methods used parallaxes and proper motions of known young stars to identify different star formation groups. While such methods are beyond the scope of this letter, we used {\it Gaia} extinctions which are part of the DR3 release to look for molecular clouds in the direction of DraChi. Molecular clouds increase the extinction of the stars behind them, thus producing breakpoints in absorption along the line of sight, therefore clouds can be found by identifying the breakpoints in absorption \citep{yan2019}.

\begin{figure}
\begin{center}
\includegraphics[width=0.45\textwidth]
{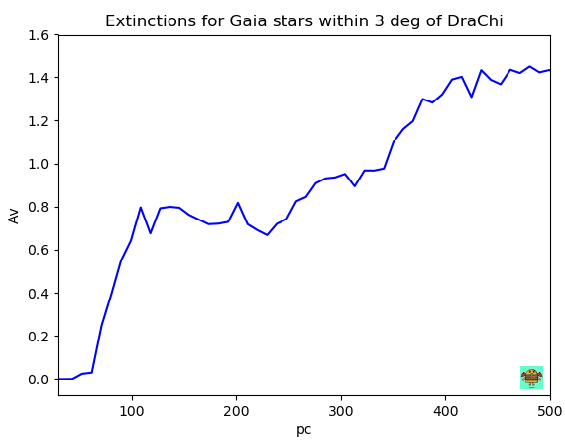}
\end{center}
\caption{Average extinctions for {\it Gaia} stars within 3$\degree$ of DraChi.}
\label{fig:extinctions}
\end{figure}

We selected stars located within 3$\degree$ distance from DraChi. The {\it Gaia} extinctions were converted to  A$_V$ using the dustapprox tool\footnote{https://github.com/mfouesneau/dustapprox}. The result is shown in Figure \ref{fig:extinctions}. The nearest cloud seems to be located between 70 and 130 pc with A$_V$ up to 0.7 and the next larger cloud between 250 and 500 pc with A$_V$ up to 1.4. 
In the modeling section we adopted the distance of 300 pc which places DraChi at the front of the second cloud. We give below some qualitative arguments for preferring this distance but we emphasize this is not a precise measurement by any means. We expect more data or more sophisticated methods as mentioned above can improve this estimate. Some of these arguments are based on the modeling results so we used our best model to run simulations for distances of 100 and 500 pc while keeping the physical parameters the same to compare the results with those for the adopted distance of 300 pc.

First we compare the disk mass ($\sim 0.2 M_\odot$) and size (1650 AU) we obtained from the best model (at 300 pc) with the values of other known disks \citep{will2011, Andrews2020}. Most known protoplanetary disks have masses $<0.1  M_\odot$ and sizes $<$ 1000 AU so our estimates are already very high compared to other disks. \citet{natta2000} investigated A-type PMS stars in particular, they show that for stars with masses $<2 M_\odot$ the disks have masses $<0.2  M_\odot$. The same disk mass limit is measured by \citet{boi2011} for stars with masses $<3 M_\odot$. For a distance of 500 pc, the disk radius is 2750 AU and the modeling gives a disk mass $> 0.5 M_\odot$. The mass ratio disk/star is in this case larger than 0.25.
Secondly, the extinction estimated from {\it Gaia} for distances corresponding to the nearest cloud in Figure \ref{fig:extinctions} are smaller than 0.7 A$_V$, and those for the next cloud go up to 1.4 A$_V$. Both our estimates based on the star spectrum and the radiative transfer modeling are 0.95 A$_V$ which is approximately the extinction that corresponds to 300 pc in Figure \ref{fig:extinctions}. The estimate from the spectrum includes both the intrinsic and the line-of-sight extinction so it is even less likely that the line-of-sight extinction is larger than 0.9 A$_V$ but we cannot rule out the a lower extinction which corresponds to the nearest cloud. We also note that the absorption estimated from the spectrum is based on the reddening and not the total extinction which is expected to be very high. The total absorbed luminosity estimated from the best model is about 2.5 L$_{\odot}$ and the estimated luminosity of the star is 11.46 L$_{\odot}$ with most of the extinction in the optical where the disk is optically thick. 
Both these arguments point against a distance as high as 500 pc.

The final argument is against a lower distance of about 100 pc, based on the star luminosity obtained from the modeling and the spectral type of the star obtained from the optical spectrum. For our 300 pc model we obtained a luminosity of 11.5 L$_{\odot}$ but for a distance of 100 pc the star luminosity obtained from the model would be 5.2 L$_{\odot}$, which is low for a HAeBe star \citep{mont2009}. We point out that these estimates of the star luminosity are based on fitting the model to the SED and in particular to the mid and far-IR which are not affected much by extinction. However, there other factors that contribute to the luminosity estimates such as disk mass or flaring.

For all these reasons we prefer the distance of $\sim$300~pc but we emphasize again that this is not a precise estimate.

\section{DraChi: the largest known protoplanetary disk}\label{sec:discussion}

DraChi is one of only three known HAeBe edge-on protoplanetary disks. As already pointed out, it shares quite a few similarities with GoHam, including a large apparent size, but DraChi is about 50\% larger. Located a distance of 250-300 pc, the $\sim2.5 M_\odot$ star in GoHam is surrounded by a 0.3 $M_\odot$ disk with an inclination of $\sim86\degree$ \citep{wood2008, buj2008}. These values are quite close to our estimates for DraChi, and we followed the radiative transfer modeling of GoHam in \citet{wood2008}, including the dust properties. We note that GoHam presents evidence for a possible protoplanet \citep{bern2015, teag2020}, showing that planet formation might be detected in such edge-on disks. 

One important difference from DraChi is the absence of an envelope as GoHam is thought to be a more evolved disk, while DraChi appears to still be surrounded by a faint and probably dissipating envelope if our interpretation of the ``fangs" is correct. Similar filaments or ``rays" are present in HV Tau C, extending out to almost double the size of the disk. They are closer to the mid-plane and straighter compared to our ``fangs". \citet{stap2003} and \citet{duch2010} give different interpretation for these filaments in HV Tau C, as part of the disk or the envelope, which is quite bright in this edge-on system. For DraChi we prefer the second option, the fangs are quite similar to the ``edges'' of the envelope in Figure~\ref{fig:image}. In addition, extended emission can be seen from the fainter lobe and additional faint filaments in the northern part of this lobe, extending perpendicular to the disk. Another bright envelope is present in the Class I disk called the Butterfly \citep{Villenave2024}, in this case the envelope dominates the scattered light emission and shows clear filaments extending out to large distances.

The third  known HAeBe edge-on protoplanetary disk is PDS 144N wth an inclination of 83$\degree$. This system has no conspicuous envelope but shows pronounced flaring at the edges of the disk, similar to what we see at the lower edge of the bright lobe of Drachi (Figure \ref{fig:image}). \citet{perr2006} was able to simulate these ``wings'' with a model that included an envelope and cavity and a dust model which included PAHs.

Evidence for grain settling was found in many edge-on disks such as HH 30 IRS \citep{wood2002} and Oph163131 \citep{Villenave2022}. It is unclear how long the grain settling takes place, but recent studies suggest most settling takes place between Class I and II \citep{Villenave2024} where we think DraChi is currently. \citet{Villenave2020} showed that some edge-on disks show evidence that large grain settling takes place at scale heights significantly smaller than the small grains. Our radiative transfer model simulated grain settling by adding a disk with large grains at a scale height six times smaller than the small grains. Constraining the grain size is very important in modeling disks, but requires observations in millimeter \citep{monsch2024}. For example, it is well known that grain size distributions and opacities have a strong impact on disk mass estimates. If large grains are included compared to ISM type dust the mass estimates are increased significantly as the gas is decoupled from the dust \citep{wood2002}. One example is HK Tau B \citep{duch2003}, where estimates of the disk mass were shown to vary by a factor of 5 depending on the grains used. 

We detected a lateral asymmetry in DraChi, similar to many other edge-on disks. \citet{Villenave2024} found that 15 out of 20 edge-on disks in their sample show clear lateral asymmetries, including GoHam and PDS 144N, and they attribute this effect to a tilted inner region of the disk. Such a disk also changes the brightness of the lobes in edge-on disks and can even lead to an inversion of the brightest lobes at specific wavelengths as they detect in mid-infrared in the Butterfly disk. We also found a significant lateral asymmetry in DraChi (panel d of Figure \ref{fig:image}) and therefore our estimate of the disk inclination in Drachi based on the brightness of the lobes could be biased.

\section{Conclusion}\label{sec:conclusion}

We have identified a new, very large protoplanetary disk from PS1 images and confirmed the consistency of available data with a disk interpretation through radiative transfer and SED modelling. The host star was estimated to be an A9 HAeBe star based on an archival optical spectrum. 
We use the PS1 image to constrain the radiative transfer model parameters and the SED to find the best model. Our data is insufficient to constrain the grain sizes and spatial distribution and we therefore employed a model used successfully on other edge-on disks including GoHam. Our model matches the observed SED reasonably well, although larger discrepancies are present in the mid-IR. The model disk has a mass of 0.2 $M_{\odot}$, an inclination angle of 82$\degree$, and its radius is 1650~AU at the adopted distance of 300~pc estimated from {\it Gaia} stellar extinctions. The low-density envelope included in the model has a 3000~AU radius and reproduces well the faint ``fangs" seen in the northern part of the disk. If this structure is indeed part of the envelope, it suggests a young system at the end of the Class I phase. We found a lateral asymmetry in the disk as detected in many edge-on disks, suggesting a tilted inner region.  The serendipitous discovery of DraChi implies that such edge-on disks not associated with known star-forming regions (GoHam is another case) are still waiting to be discovered.

\begin{acknowledgments}
We are grateful to the referee for very helpful suggestions that greatly improved our paper.
The Pan-STARRS1 Surveys (PS1) and the PS1 public science archive have been made possible through contributions by the Institute for Astronomy, the University of Hawaii, the Pan-STARRS Project Office, the Max-Planck Society and its participating institutes, the Max Planck Institute for Astronomy, Heidelberg and the Max Planck Institute for Extraterrestrial Physics, Garching, The Johns Hopkins University, Durham University, the University of Edinburgh, the Queen's University Belfast, the Harvard-Smithsonian Center for Astrophysics, the Las Cumbres Observatory Global Telescope Network Incorporated, the National Central University of Taiwan, the Space Telescope Science Institute, the National Aeronautics and Space Administration under Grant No. NNX08AR22G issued through the Planetary Science Division of the NASA Science Mission Directorate, the National Science Foundation Grant No. AST-1238877, the University of Maryland, Eotvos Lorand University (ELTE), the Los Alamos National Laboratory, and the Gordon and Betty Moore Foundation.
\end{acknowledgments}

\begin{deluxetable*}{ccccc}[hb!]
\tablenum{1}
\tablecaption{Photometry\label{tbl:photometry}}
\tablewidth{0pt}
\tablehead{
\colhead{\textbf{Wavelength ($\mu$m)}} & \colhead{\textbf{Flux (mJy)}} &
\colhead{\textbf{Source}} &
\colhead{\textbf{Instrument}} & \colhead{\textbf{Angular Resolution($"$)}} 
}
\startdata
0.23 & 0.14 $\pm$ 0.1 & \cite{bianchi2011} & GALEX & 5.3\\\hline
0.48 & 6.46 $\pm$ 0.20 & This Work & PS1 & $\sim$1.0 \\
0.62 & 8.52 $\pm$ 0.39 & This Work & PS1 & $\sim$1.0 \\
0.75 & 9.71 $\pm$ 0.37 & This Work & PS1 & $\sim$1.0 \\
0.87 & 10.67 $\pm$ 0.50 & This Work & PS1 & $\sim$1.0 \\
0.96 & 12.41 $\pm$ 0.80 & This Work & PS1 & $\sim$1.0 \\\hline
1.23 & 14.84 $\pm$ 0.45 & \cite{skrutskie2006} & 2MASS & $\sim$5\\
1.66 & 21.1 $\pm$ 0.60 & \cite{skrutskie2006} & 2MASS & $\sim$5\\
2.16 & 39.2 $\pm$ 0.76 & \cite{skrutskie2006} & 2MASS & $\sim$5\\\hline
3.4 & 57.3 $\pm$ 1.21 & \cite{cutri2014} & ALLWISE & 6.1\\
4.6 & 116 $\pm$ 2.14 & \cite{cutri2014} & ALLWISE & 6.4\\
12 & 310 $\pm$ 3.10 & \cite{cutri2014} & ALLWISE & 6.5\\
22 & 318 $\pm$ 5.27 & \cite{cutri2014} & ALLWISE & 12.0\\\hline
12 & 315 $\pm$ 25 & \cite{helou1988} & IRAS & $\sim$30\\
25 & 407 $\pm$ 33 & \cite{helou1988} & IRAS & $\sim$30\\
60 & 10700 $\pm$ 86 & \cite{helou1988} & IRAS & $\sim$60\\
100 & 14400 $\pm$ 115 & \cite{helou1988} & IRAS & $\sim$120\\\hline
8.6 & 404 $\pm$ 12 & \cite{ishi2010} & AKARI/IRC & 5.5\\
18.4 & 220 $\pm$ 18 & \cite{ishi2010} & AKARI/IRC & 5.5\\
65 & 10840 $\pm$ 570 & \cite{kaw2007} & AKARI/FIS & 37\\
90 & 15490 $\pm$ 276 & \cite{kaw2007} & AKARI/FIS & 39\\
140 & 15440 $\pm$ 1350 & \cite{kaw2007} & AKARI/FIS & 58\\
160 & 12750 $\pm$ 842 & \cite{kaw2007} & AKARI/FIS & 61
\enddata
\tablecomments{Photometry presented in this paper from PS1 images and archival data available for this object.}
\end{deluxetable*}

\bibliographystyle{aasjournal}

\begin{thebibliography}{}

\bibitem[ALMA Partnership et al.(2015)]{ALMAPartnership2015} ALMA Partnership, Brogan, C.~L., P{\'e}rez, L.~M., et al.\ 2015, \apjl, 808, L3. doi:10.1088/2041-8205/808/1/L3

\bibitem[Andrews(2020)]{Andrews2020} Andrews, S.~M.\ 2020, \araa, 58, 483. doi:10.1146/annurev-astro-031220-010302

\bibitem[Angelo et al.(2023)]{Angelo2023} Angelo, I., Duchene, G., Stapelfeldt, K., et al.\ 2023, \apj, 945, 130. doi:10.3847/1538-4357/acbb01

\bibitem[Appenzeller et al.(2005)]{app2005} Appenzeller, I., Bertout, C., \& Stahl, O.\ 2005, \aap, 434, 1005. doi:10.1051/0004-6361:20042217

\bibitem[Bern{\'e} et al.(2015)]{bern2015} Bern{\'e}, O., Fuente, A., Pantin, E., et al.\ 2015, \aap, 578, L8. doi:10.1051/0004-6361/201526041

\bibitem[Bianchi et al.(2011)]{bianchi2011} Bianchi, L., Herald, J., Efremova, B., et al.\ 2011, \apss, 335, 161 

\bibitem[Binkert(2023)]{Binkert2023} Binkert, F.\ 2023, \mnras, 525, 4299. doi:10.1093/mnras/stad2471

\bibitem[Boissier et al.(2011)]{boi2011} Boissier, J., Alonso-Albi, T., Fuente, A., et al.\ 2011, \aap, 531, A50. doi:10.1051/0004-6361/201016306

\bibitem[Bujarrabal et al.(2008)]{buj2008} Bujarrabal, V., Young, K., \& Fong, D.\ 2008, \aap, 483, 839. doi:10.1051/0004-6361:20079273


\bibitem[Chambers et al.(2016)]{cha2016} Chambers, K.~C., Magnier, E.~A., Metcalfe, N., et al.\ 2016, arXiv:1612.05560. doi:10.48550/arXiv.1612.05560

\bibitem[Cox(2000)]{cox2000} Cox, A.~N.\ 2000, Allen's astrophysical quantities, 4th ed. Publisher: New York: AIP Press; Springer, 2000. Editedy by Arthur N. Cox.  ISBN: 0387987460

\bibitem[Cutri et al.(2013)]{cutri2013} Cutri, R.~M., Wright, E.~L., Conrow, T., et al.\ 2013, Explanatory Supplement to the AllWISE Data Release Products, by R. M. Cutri et al.

\bibitem[Cutri et al.(2014)]{cutri2014} Cutri, R. M., et al. 2014, VizieR On-line Data Catalog, 2311, 0

\bibitem[Draine \& Li(2007)]{draine2007} Draine, B. T., \& Li, A. 2007, \apj, 657, 810. doi:10.1086/511055

\bibitem[Duch{\^e}ne et al.(2003)]{duch2003} Duch{\^e}ne, G., M{\'e}nard, F., Stapelfeldt, K., et al.\ 2003, \aap, 400, 559. doi:10.1051/0004-6361:20021906

\bibitem[Duch{\^e}ne et al.(2010)]{duch2010} Duch{\^e}ne, G., McCabe, C., Pinte, C., et al.\ 2010, \apj, 712, 112. doi:10.1088/0004-637X/712/1/112

\bibitem[Duch{\^e}ne et al.(2024)]{duch2024} Duch{\^e}ne, G., M{\'e}nard, F., Stapelfeldt, K.~R., et al.\ 2024, \aj, 167, 77. doi:10.3847/1538-3881/acf9a7

\bibitem[Dullemond \& Dominik(2005)]{Dullemond2005} Dullemond, C.~P. \& Dominik, C.\ 2005, \aap, 434, 971. doi:10.1051/0004-6361:20042080

\bibitem[Dutrey et al.(2017)]{Dutrey2017} Dutrey, A., Guilloteau, S., Pi{\'e}tu, V., et al.\ 2017, \aap, 607, A130. doi:10.1051/0004-6361/201730645

\bibitem[Espaillat et al.(2011)]{esp2011} Espaillat, C., Furlan, E., D'Alessio, P., et al.\ 2011, \apj, 728, 49. doi:10.1088/0004-637X/728/1/49

\bibitem[Flores et al.(2021)]{flores2021} Flores, C., Duch{\^e}ne, G., Wolff, S., et al.\ 2021, \aj, 161, 239. doi:10.3847/1538-3881/abeb1e

\bibitem[Goldreich \& Ward(1973)]{Goldreich1973} Goldreich, P. \& Ward, W.~R.\ 1973, \apj, 183, 1051. doi:10.1086/152291

\bibitem[Helou et al. (1988)]{helou1988} Helou, G., \& Walker, D. W., eds. 1988, IRAS Catalogs: The Small Scale Structure Catalog (Washington, DC: GPO)

\bibitem[Herbig(1960)]{her1960} Herbig, G.~H.\ 1960, \apjs, 4, 337. doi:10.1086/190050


\bibitem[Ishihara et al.(2010)]{ishi2010} Ishihara, D., Onaka, T., Kataza, H., et al.\ 2010, \aap, 514, A1

\bibitem[Jarrett et al.(2000)]{jarr2000} Jarrett, T.~H., Chester, T., Cutri, R., et al.\ 2000, \aj, 119, 2498. doi:10.1086/301330

\bibitem[Kawada et al.(2007)]{kaw2007} Kawada, M., Baba, H., Barthel, P.~D., et al.\ 2007, \pasj, 59, S389

\bibitem[Kesseli et al.(2017)]{kess2017} Kesseli, A.~Y., West, A.~A., Veyette, M., et al.\ 2017, \apjs, 230, 16

\bibitem[Kim et al.(1994)]{kim1994} Kim, S.-H., Martin, P. G., \& Hendry, P. D. 1994, ApJ, 422, 164. doi:10.1086/173714

\bibitem[Le D{\^u} et al.(2022)]{ledu2022} Le D{\^u}, P., Mulato, L., Parker, Q.~A., et al.\ 2022, \aap, 666, A152


\bibitem[Miotello et al.(2014)]{miotello2014} Miotello, A., Testi, L., Lodato, G., et al. 2014, \aap 567, A32


\bibitem[Montesinos et al.(2009)]{mont2009} Montesinos, B., Eiroa, C., Mora, A., et al.\ 2009, \aap, 495, 901. doi:10.1051/0004-6361:200810623

\bibitem[Monsch et al.(2024)]{monsch2024} Monsch, K., Lovell, J.~B., Berghea, C.~T., et al.\ 2024, arXiv:2402.01941. doi:10.48550/arXiv.2402.01941


\bibitem[Natta et al.(2000)]{natta2000} Natta, A., Grinin, V., \& Mannings, V.\ 2000, Protostars and Planets IV, 559


\bibitem[Olano et al.(2006)]{olano2006} Olano, C.~A., Meschin, P.~I., \& Niemela, V.~S.\ 2006, \mnras, 369, 867. doi:10.1111/j.1365-2966.2006.10343.x

\bibitem[Palla \& Stahler(1999)]{palla1999} Palla, F. \& Stahler, S.~W.\ 1999, \apj, 525, 772. doi:10.1086/307928

\bibitem[Perrin et al.(2006)]{perr2006} Perrin, M.~D., Duch{\^e}ne, G., Kalas, P., et al.\ 2006, \apj, 645, 1272. doi:10.1086/504510

\bibitem[Pikhartova et al.(2021)]{Pik2021} Pikhartova, M., Long, Z.~C., Assani, K.~D., et al.\ 2021, \apj, 919, 64. doi:10.3847/1538-4357/ac03af


\bibitem[Pinte et al.(2016)]{Pinte2016} Pinte, C., Dent, W.~R.~F., M{\'e}nard, F., et al.\ 2016, \apj, 816, 25. doi:10.3847/0004-637X/816/1/25


\bibitem[Robitaille et al.(2006)]{rob2006} Robitaille, T.~P., Whitney, B.~A., Indebetouw, R., et al.\ 2006, \apjs, 167, 256. doi:10.1086/508424

\bibitem[Ruge et al.(2016)]{Ruge2016} Ruge, J.~P., Flock, M., Wolf, S., et al.\ 2016, \aap, 590, A17. doi:10.1051/0004-6361/201526616

\bibitem[Ruiz et al.(1987)]{ruiz1987} Ruiz, M.~T., Blanco, V., Maza, J., et al.\ 1987, \apjl, 316, L21

\bibitem[Roulston et al.(2020)]{roul2020} Roulston, B.~R., Green, P.~J., \& Kesseli, A.~Y.\ 2020, \apjs, 249, 34

\bibitem[Secrest et al. (2015)]{secrest2015} Secrest, N.~J., Dudik, R.~P., Dorland, B.~N., et al.\ 2015, ApJ, 221, 12

\bibitem[Sharma et al.(2022)]{sharma2022} Sharma, E., Maheswar, G., \& Dib, S.\ 2022, \aap, 658, A55. doi:10.1051/0004-6361/202140495

\bibitem[Stapelfeldt et al.(2003)]{stap2003} Stapelfeldt, K.~R., M{\'e}nard, F., Watson, A.~M., et al.\ 2003, \apj, 589, 410. doi:10.1086/374374

\bibitem[Szil{\'a}gyi et al.(2021)]{szi2021} Szil{\'a}gyi, M., Kun, M., \& {\'A}brah{\'a}m, P.\ 2021, \mnras, 505, 5164. doi:10.1093/mnras/stab1496

\bibitem[Skrutskie et al.(2006)]{skrutskie2006} Skrutskie, M.~F., Cutri, R.~M., Stiening, R., et al.\ 2006, \aj, 131, 1163 


\bibitem[Teague et al.(2020)]{teag2020} Teague, R., Jankovic, M.~R., Haworth, T.~J., et al.\ 2020, \mnras, 495, 451. doi:10.1093/mnras/staa1167

\bibitem[T{\'o}th et al.(2014)]{toth2014} T{\'o}th, L.~V., Marton, G., Zahorecz, S., et al.\ 2014, \pasj, 66, 17. doi:10.1093/pasj/pst017

\bibitem[Ulrich(1976)]{ulr1976} Ulrich, R.~K.\ 1976, \apj, 210, 377. doi:10.1086/154840

\bibitem[Villenave et al.(2020)]{Villenave2020} Villenave, M., M{\'e}nard, F., Dent, W.~R.~F., et al.\ 2020, \aap, 642, A164. doi:10.1051/0004-6361/202038087

\bibitem[Villenave et al.(2022)]{Villenave2022} Villenave, M., Stapelfeldt, K.~R., Duch{\^e}ne, G., et al.\ 2022, \apj, 930, 11. doi:10.3847/1538-4357/ac5fae

\bibitem[Villenave et al.(2024)]{Villenave2024} Villenave, M., Stapelfeldt, K.~R., Duch{\^e}ne, G., et al.\ 2024, \apj, 961, 95. doi:10.3847/1538-4357/ad0c4b


\bibitem[Voelk et al.(1980)]{Voelk1980} Voelk, H.~J., Jones, F.~C., Morfill, G.~E., et al.\ 1980, \aap, 85, 316

\bibitem[Waters et al.(2020)]{wat2020} Waters, C.~Z., Magnier, E.~A., Price, P.~A., et al.\ 2020, \apjs, 251, 4

\bibitem[Weidenschilling(1977)]{Weidenschilling1977} Weidenschilling, S.~J.\ 1977, \mnras, 180, 57. doi:10.1093/mnras/180.2.57

\bibitem[Whitney et al.(2013)]{whitney2013} Whitney, B.~A., Robitaille T.~P., Bjorkman, J.~E., et al.\ 2013, \apjs, 207, 2

\bibitem[Williams \& Cieza(2011)]{will2011} Williams, J.~P. \& Cieza, L.~A.\ 2011, \araa, 49, 67. doi:10.1146/annurev-astro-081710-102548

\bibitem[Winn \& Fabrycky(2015)]{Winn2015} Winn, J.~N. \& Fabrycky, D.~C.\ 2015, \araa, 53, 409. doi:10.1146/annurev-astro-082214-122246


\bibitem[Wolff et al.(2021)]{wolff2021} Wolff, S.~G., Duch{\^e}ne, G., Stapelfeldt, K.~R., et al.\ 2021, \aj, 161, 238

\bibitem[Wood et al.(2002)]{wood2002} Wood, K., Wolff, M. J., Bjorkman, J. E., \& Whitney, B. 2002, ApJ, 564, 887
doi:10.1086/324285

\bibitem[Wood et al.(2008)]{wood2008} Wood, K., Whitney, B.~A., Robitaille, T., et al.\ 2008, \apj, 688, 1118. doi:10.1086/592185

\bibitem[Wouterloot et al.(1993)]{wouterloot1993} Wouterloot, J.~G.~A., Brand, J., \& Fiegle, K.\ 1993, \aaps, 98, 589 

\bibitem[Yan et al.(2019)]{yan2019} Yan, Q.-Z., Zhang, B., Xu, Y., et al.\ 2019, \aap, 624, A6. doi:10.1051/0004-6361/201834337


\bibitem[Zacharias et al. (2015)]{zacharias2015} Zacharias, N., Finch, C., Subasavage, J., et al.\ 2015, AJ, 150, 101

\bibitem[STScI (2022)]{ps1dr2} STScI\ 2022, Pan-STARRS1 DR2 Catalog, STScI/MAST, doi=10.17909/s0zg-jx37

\end{thebibliography}

\end{document}